\documentclass[journal]{IEEEtran}

\usepackage{amsmath,amssymb,amsfonts}
\usepackage{algorithmic}
\usepackage{graphicx}
\usepackage{booktabs}
\usepackage{multirow}
\usepackage{array}
\usepackage{xcolor}
\usepackage{url}
\usepackage{cite}
\usepackage{amsthm}

\newtheorem{definition}{Definition}

\usepackage{hyperref}
\hypersetup{colorlinks=true, linkcolor=blue, citecolor=blue, urlcolor=blue}

\begin{document}

\title{Cross-Platform Domain Adaptation for Multi-Modal\\
MOOC Learner Satisfaction Prediction}

\author{
  \IEEEauthorblockN{Jakub Kowalski\IEEEauthorrefmark{1},
                    Magdalena Piotrowska\IEEEauthorrefmark{2}}
}


\maketitle

\begin{abstract}
Learner satisfaction prediction from MOOC reviews and behavioral
logs is valuable for course quality improvement and platform
operations. In practice, models trained on one platform degrade
significantly when deployed on another due to domain shift in review
style, learner population, behavioral logging schemas, and
platform-specific rating norms. We study \textbf{cross-platform
domain adaptation} for multi-modal MOOC satisfaction prediction
under limited or absent target-platform labels. We propose
\textbf{ADAPT-MS}, a platform-adaptive framework that (i) encodes
review text with a frozen LLM encoder and behavioral traces with a
canonical-vocabulary MLP, (ii) aligns cross-platform representations
via domain-adversarial training with gradient reversal, (iii)
corrects platform-specific rating bias through a latent-variable
calibration layer, and (iv) handles missing behavioral modalities via
gated fusion with modality dropout. Experiments on a multi-platform
MOOC dataset spanning three major platforms demonstrate that
ADAPT-MS achieves target-platform RMSE of 0.66 in the unsupervised
setting (zero labeled target samples) and 0.60 with 1000 labeled
target samples, outperforming strong baselines including naive
pooling, domain-adversarial alignment without calibration, and
full fine-tuning. Ablation studies confirm the independent
contribution of each component, and few-shot adaptation curves
demonstrate stable improvement even with as few as 50 labeled
target samples.
\end{abstract}


\section{Introduction}
\label{sec:intro}

\IEEEPARstart{M}{assive} open online courses (MOOCs) have scaled
globally, with learners enrolling across diverse platforms such as
Coursera, edX, and XuetangX \cite{reich2019mooc, shah2020mooc, chi2024active}.
These platforms accumulate rich feedback signals---star ratings,
textual reviews, and behavioral interaction logs---that can be
mined to predict and improve learner satisfaction
\cite{qi2021evaluating, hew2020predict, almatrafi2019forums}.
Recent advances in pretrained language models \cite{devlin2019bert,
liu2019roberta} and multi-modal fusion have enabled accurate
satisfaction prediction within a single platform.

However, real-world deployment requires models that generalize
across platforms. Platform heterogeneity manifests along at least
four axes:
\begin{enumerate}
  \item \emph{Linguistic style shift}: review text from different
    platforms differs in vocabulary, formality, and cultural
    expression of sentiment.
  \item \emph{Population and course mix shift}: platforms attract
    different learner demographics and subject-area distributions.
  \item \emph{Behavioral logging schema shift}: the same interaction
    (e.g., ``video view'') may be defined differently or logged at
    different granularities.
  \item \emph{Rating scale semantics drift}: learners on some
    platforms reserve 5 stars for perfection, while others use 5
    stars as the default positive rating, creating systematic bias
    in predicted satisfaction levels.
\end{enumerate}
As a result, a model trained on a source platform $\mathcal{S}$
may exhibit substantial degradation when deployed on a target
platform $\mathcal{T}$, even when the underlying satisfaction
mechanisms are similar.

Existing work on MOOC satisfaction prediction largely ignores
cross-platform generalization, evaluating models on held-out samples
from the same platform used for training \cite{qi2021evaluating,
onan2020mooc, hew2020predict}. Classical domain adaptation methods
\cite{pan2010transfer, pan2011tca, ganin2017dann} address distribution
shift but do not model label semantics drift or multi-modal missingness
jointly.

To bridge this gap, we propose \textbf{ADAPT-MS}
(\underline{Adap}tive \underline{D}omain-\underline{T}ransfer
\underline{M}ulti-modal \underline{S}atisfaction predictor), a
unified framework illustrated in Fig.~\ref{fig:arch}. ADAPT-MS
fuses LLM text embeddings and canonical behavioral features,
aligns cross-platform representations via a gradient-reversal
domain discriminator, corrects platform-specific rating bias with
a latent-variable calibration layer, and maintains robustness to
missing behavioral modalities through gated fusion and modality
dropout.

\textbf{Main contributions.} This paper makes the following
contributions:
\begin{enumerate}
  \item We identify and quantify four key sources of cross-platform
    shift for MOOC satisfaction prediction: feature distribution
    shift, behavioral logging gaps, population mix shift, and rating
    scale semantics drift.
  \item We propose ADAPT-MS, a domain-adaptive multi-modal framework
    that jointly addresses representation alignment and label
    calibration under a missing-modality-robust fusion architecture.
  \item We design a rigorous cross-platform evaluation protocol
    (unsupervised, few-shot, and pairwise transfer) and demonstrate
    consistent improvements over strong baselines with detailed
    ablations and sensitivity analyses.
\end{enumerate}

The remainder of this paper is organized as follows.
Section~\ref{sec:related} reviews related work.
Section~\ref{sec:problem} defines the cross-platform adaptation
problem. Section~\ref{sec:method} presents ADAPT-MS in detail.
Section~\ref{sec:exp} reports experimental results.
Section~\ref{sec:conclusion} concludes the paper.

\section{Related Work}
\label{sec:related}

\subsection{MOOC Satisfaction Prediction and Review Mining}
Automated satisfaction mining for MOOCs has been studied using
lexicon-based sentiment analysis, aspect-level opinion extraction,
and supervised learning on review text \cite{qi2021evaluating,
onan2020mooc, kastrati2020aspect}. Hew \emph{et al.} showed that
combining sentiment features with gradient-boosted classifiers
achieves competitive satisfaction prediction \cite{hew2020predict}.
Behavioral traces have also been used to augment text-based
models, exploiting the complementarity of interaction logs and
textual feedback \cite{hew2016engagement, peng2020behavior}.
However, virtually all existing work evaluates within a single
platform, leaving cross-platform generalization unaddressed.

\subsection{Transfer Learning and Domain Adaptation}
Pan and Yang surveyed transfer learning and highlighted covariate
shift as the dominant challenge for cross-domain deployment
\cite{pan2010transfer}. Transfer Component Analysis (TCA) aligns
feature distributions via kernel-based moment matching
\cite{pan2011tca}. Domain-adversarial neural networks (DANN) by
Ganin \emph{et al.} introduced gradient reversal to learn
domain-invariant features \cite{ganin2017dann}, and subsequent
deep domain adaptation surveys have extended these ideas to a wide
range of architectures \cite{wang2018deepda}. In educational data
mining, transfer learning has been applied to knowledge tracing
and dropout prediction \cite{pan2010transfer}, but cross-platform
satisfaction prediction with label calibration has not been
systematically studied.

\subsection{Label Shift and Output Calibration}
Label shift---where the marginal label distribution $p(y)$ differs
between source and target---is distinct from covariate shift
\cite{lipton2018detecting}. Calibration methods such as temperature
scaling and Platt scaling correct prediction confidence
\cite{platt1999probabilistic}. Rating scale differences in
satisfaction surveys are a structured form of label shift that is
amenable to affine calibration. We leverage this structure to
design a lightweight but effective calibration layer that can be
adapted with very few labeled target samples.

\subsection{Multi-Modal Learning with Missing Modalities}
Multi-modal fusion \cite{baltrusaitis2019multimodal} typically
assumes all modalities are available at both training and inference.
Missing modality robustness has been studied in vision-language
models and clinical prediction \cite{ma2021smil}, where modality
dropout and gating mechanisms provide effective solutions. In MOOC
settings, behavioral logs are frequently absent on the target
platform or incompatible with the source schema, making missing-
modality robustness essential for cross-platform deployment.

\subsection{LLM Representations for Short Text}
BERT \cite{devlin2019bert} and RoBERTa \cite{liu2019roberta}
provide powerful contextual representations for short, noisy text.
Freezing the pretrained encoder and training only a task-specific
head is competitive in low-resource settings and reduces the risk
of source-platform overfitting, which is particularly important
when target data are scarce.

\section{Problem Formulation}
\label{sec:problem}

Let $\mathcal{P} = \{1, \dots, P\}$ index a set of MOOC platforms.
Each enrollment instance from platform $p$ is a tuple
$(r_i, \mathbf{b}_i, y_i, p_i)$, where $r_i$ is the review text,
$\mathbf{b}_i \in \mathbb{R}^{d_b}$ is the behavioral feature
vector, $y_i \in [1,5]$ is the satisfaction label, and
$p_i \in \mathcal{P}$ is the platform indicator.

\begin{definition}[Cross-Platform Shift]
Let $\mathcal{D}_p = \{(r_i, \mathbf{b}_i, y_i)\}$ denote the
distribution on platform $p$. Cross-platform shift is the joint
discrepancy $\mathcal{D}_{\mathcal{S}} \neq \mathcal{D}_{\mathcal{T}}$,
encompassing feature distribution shift in $(r_i, \mathbf{b}_i)$,
missing-modality patterns, and label semantics shift in $y_i$.
\end{definition}

\begin{definition}[Cross-Platform Adaptation]
Given labeled source data $\{(r_i, \mathbf{b}_i, y_i)\}$ from
platforms $\mathcal{S} \subset \mathcal{P}$ and (possibly unlabeled)
target data $\{(r_j, \mathbf{b}_j)\}$ from target platform
$\mathcal{T} \notin \mathcal{S}$, learn a predictor $f$ that
minimizes expected target-platform error:
\begin{equation}
  \min_f\; \mathbb{E}_{(r,\mathbf{b},y)\sim\mathcal{D}_{\mathcal{T}}}
  \bigl[\ell\!\left(f(r,\mathbf{b}),\, y\right)\bigr],
  \label{eq:objective}
\end{equation}
using only the source labels and unlabeled (or few-labeled) target
instances.
\end{definition}

We evaluate three adaptation regimes: (i) \textit{unsupervised}
($k=0$ labeled target samples), (ii) \textit{few-shot}
($k \in \{50, 200, 1000\}$), and (iii) \textit{pairwise transfer}
(train on one source platform, test on another).

\section{Proposed Method: ADAPT-MS}
\label{sec:method}


\subsection{Multi-Modal Encoder with Canonical Behavior Features}

\subsubsection{Text Encoder}
Review text $r_i$ is encoded using a frozen pretrained LLM
(RoBERTa-base \cite{liu2019roberta}). We extract the
\texttt{[CLS]} token representation:
\begin{equation}
  \mathbf{h}_i = \mathrm{PLM}(r_i)_{\texttt{[CLS]}}
               \;\in\mathbb{R}^{d_{\mathrm{llm}}}.
  \label{eq:text_enc}
\end{equation}
Freezing the encoder reduces the risk of overfitting to
source-platform vocabulary while retaining general linguistic
representations suitable for cross-platform transfer.

\subsubsection{Canonical Behavioral Features}
Because platforms log different low-level event types, we map
raw behavioral logs to a platform-agnostic canonical feature
vector:
\begin{equation}
  \mathbf{b}_i^{\mathrm{can}} = \bigl[
    v_i,\; q_i,\; f_i^r,\; f_i^w,\; d_i,\; \rho_i
  \bigr] \;\in\mathbb{R}^{d_b},
  \label{eq:behavior}
\end{equation}
where $v_i$ = minutes watched, $q_i$ = quiz attempts,
$f_i^r$ = forum reads, $f_i^w$ = forum posts,
$d_i$ = active days, and $\rho_i$ = video rewatch rate.
These six aggregates are computable from any standard MOOC
logging schema. A two-layer MLP $\phi$ with batch normalization
maps $\mathbf{b}_i^{\mathrm{can}}$ to a $d$-dimensional
behavioral embedding $\mathbf{g}_i = \phi(\mathbf{b}_i^{\mathrm{can}})$.

\subsubsection{Gated Fusion with Modality Dropout}
Let $m_i \in \{0,1\}$ indicate whether behavioral logs are
available for instance $i$. A scalar gate controls the behavioral
contribution:
\begin{equation}
  \alpha_i = \sigma\!\left(\mathbf{w}_g^\top
    [\mathbf{h}_i;\, m_i] + b_g\right),
  \label{eq:gate}
\end{equation}
and the fused representation is
\begin{equation}
  \mathbf{z}_i = \bigl[\mathbf{h}_i;\; \alpha_i\,\mathbf{g}_i\bigr].
  \label{eq:fusion}
\end{equation}
During training, we apply modality dropout by randomly zeroing
the behavioral modality with probability $p_{\mathrm{mod}}=0.3$,
forcing the model to rely on text alone for a fraction of
training steps. This prevents degenerate behavior when
$\mathbf{g}_i$ is unavailable at inference on the target platform.

\subsection{Domain-Adversarial Representation Alignment}

To encourage platform-invariant representations, we train a
domain discriminator $D_\psi$ to predict the platform label
$p_i \in \{1,\dots,P\}$ from the shared representation
$\mathbf{z}_i$. A gradient reversal layer (GRL)
\cite{ganin2017dann} is inserted between the encoder and the
discriminator, which multiplies gradients by $-\lambda$ during
backpropagation. The composite objective is:
\begin{equation}
  \mathcal{L}_{\mathrm{total}}
    = \mathcal{L}_{\mathrm{task}}(f(\mathbf{z}_i), y_i)
      - \lambda\,\mathcal{L}_{\mathrm{dom}}(D_\psi(\mathbf{z}_i), p_i),
  \label{eq:total_loss}
\end{equation}
where $\mathcal{L}_{\mathrm{task}}$ is the mean squared error
(MSE) on labeled source data, and $\mathcal{L}_{\mathrm{dom}}$
is the cross-entropy for the platform classification task.
The alignment weight $\lambda$ is treated as a hyperparameter
tuned on the validation split.

\textbf{Theoretical motivation.} By the H-divergence bound from
domain adaptation theory \cite{ben2010theory}, minimizing the
source task loss and the domain divergence simultaneously provides
an upper bound on the target task loss under mild assumptions on
the hypothesis class.

\subsection{Platform-Specific Rating Calibration}

Even after representation alignment, label semantics drift
causes systematic prediction bias. We model a latent satisfaction
score $s_i = f(\mathbf{z}_i)$ and a platform-specific affine
calibration layer:
\begin{equation}
  \hat{y}_i = c_{p_i}(s_i) = a_{p_i}\,s_i + b_{p_i},
  \label{eq:calibration}
\end{equation}
where $(a_p, b_p)$ are platform-specific scale and bias
parameters.

\subsubsection{Supervised Calibration (labeled target)}
When $k > 0$ labeled target samples are available, we optimize
$(a_{\mathcal{T}}, b_{\mathcal{T}})$ jointly with a small
learning rate while keeping the shared encoder frozen.

\subsubsection{Unsupervised Calibration (zero labels)}
When no target labels exist, we estimate calibration parameters
via distribution matching. Let $\bar{s}$ and $\sigma_s$ denote
the mean and standard deviation of source predictions on the
target domain, and let $\bar{y}_{\mathcal{T}}^{\mathrm{hist}}$
and $\sigma_y^{\mathcal{T}}$ denote the historical target
rating statistics (obtainable from platform metadata or audited
summary statistics). Then:
\begin{align}
  a_{\mathcal{T}} &= \frac{\sigma_y^{\mathcal{T}}}{\sigma_s},
  \label{eq:calib_a} \\
  b_{\mathcal{T}} &= \bar{y}_{\mathcal{T}}^{\mathrm{hist}}
                     - a_{\mathcal{T}}\,\bar{s}.
  \label{eq:calib_b}
\end{align}
This moment-matching approach requires only the first two
moments of the target rating distribution, which can often be
obtained from platform dashboards or a small audited sample
without individual labels.

Table~\ref{tab:notation} summarizes the key notation.

\begin{table}[!t]
  \caption{Summary of Key Notation}
  \label{tab:notation}
  \centering
  \renewcommand{\arraystretch}{1.25}
  \begin{tabular}{cl}
    \toprule
    Symbol & Meaning \\
    \midrule
    $\mathcal{S}$, $\mathcal{T}$ & Source and target platforms \\
    $r_i$ & Review text of instance $i$ \\
    $\mathbf{b}_i^{\mathrm{can}}$ & Canonical behavioral feature vector \\
    $\mathbf{h}_i$ & LLM text embedding \\
    $\mathbf{g}_i$ & Behavioral MLP embedding \\
    $\alpha_i$ & Modality gate scalar \\
    $\mathbf{z}_i$ & Fused multi-modal representation \\
    $s_i$ & Latent satisfaction score \\
    $\hat{y}_i$ & Calibrated satisfaction prediction \\
    $(a_p, b_p)$ & Platform-specific calibration parameters \\
    $\lambda$ & Domain alignment weight \\
    $m_i$ & Behavioral modality availability indicator \\
    \bottomrule
  \end{tabular}
\end{table}

\section{Experiments}
\label{sec:exp}

\subsection{Dataset and Platform Statistics}

We conduct experiments on a multi-platform MOOC dataset covering
three major platforms (denoted A, B, and C). Each platform
contributes review text, star ratings, and partial behavioral
interaction logs. Table~\ref{tab:platform_stats} summarizes
platform-level shift diagnostics.

\begin{table}[!t]
  \caption{Cross-Platform Shift Diagnostics}
  \label{tab:platform_stats}
  \centering
  \renewcommand{\arraystretch}{1.2}
  \begin{tabular}{lccc}
    \toprule
    \textbf{Platform}
      & \textbf{Mean Rating}
      & \textbf{Avg.\ Review Length}
      & \textbf{Missing Behav.} \\
    \midrule
    A (source) & 4.35 $\pm$ 0.72 & 28 tokens & 5\%  \\
    B (source) & 4.10 $\pm$ 0.85 & 35 tokens & 12\% \\
    C (target) & 4.70 $\pm$ 0.55 & 18 tokens & 40\% \\
    \bottomrule
  \end{tabular}
\end{table}

The pronounced differences in mean rating (4.10 on B vs.\ 4.70 on C)
and missing behavioral log rates (5\% on A vs.\ 40\% on C)
motivate both the calibration and gated fusion components of
ADAPT-MS. Overall, the dataset comprises 480,000 enrollments,
95M behavioral events, and 1.8M review snippets across all platforms.

\subsection{Evaluation Protocols}

We evaluate under three protocols:
\begin{itemize}
  \item \textbf{Unsupervised transfer}: Train on A+B, test on C
    with zero labeled target samples ($k=0$).
  \item \textbf{Few-shot adaptation}: Same split, with
    $k\in\{50,200,1000\}$ labeled target samples used to
    update calibration parameters only.
  \item \textbf{Pairwise transfer}: Train on A, test on B, and
    vice versa, to assess sensitivity to specific platform pairs.
\end{itemize}
All experiments use time-based splits within each platform
(earliest 70\% for training, next 15\% for validation, most
recent 15\% for testing) to prevent temporal leakage.

\subsection{Baselines}

We compare ADAPT-MS against the following baselines:
\begin{itemize}
  \item \textbf{Source-Only}: Train on source platforms with no
    adaptation.
  \item \textbf{Pool-NoAdapt}: Pool all source data without
    domain labels; no adaptation.
  \item \textbf{Fine-Tune}: Initialize from source model and
    fine-tune all parameters on $k$ labeled target samples.
  \item \textbf{DANN-Only}: Domain-adversarial alignment
    \cite{ganin2017dann} without calibration.
  \item \textbf{ADAPT-MS (Ours)}: Full framework.
\end{itemize}

\subsection{Implementation Details}

ADAPT-MS uses RoBERTa-base \cite{liu2019roberta} (frozen) as the
text encoder ($d_{\mathrm{llm}}=768$, projected to $d=256$).
The behavioral MLP has two hidden layers of size 128 with ReLU
and batch normalization. The fusion hidden size is $d_f=512$.
The domain discriminator is a two-layer MLP with gradient reversal.
We train with the Adam optimizer \cite{kingma2015adam}, learning
rate $3\times10^{-4}$, batch size 256, and early stopping on
validation RMSE. The alignment weight $\lambda$ is tuned over
$\{0.0, 0.1, 0.3, 0.5, 1.0\}$. For few-shot adaptation, only
the calibration parameters $(a_{\mathcal{T}}, b_{\mathcal{T}})$
and the gating network are updated, keeping the shared encoder
frozen for stability. All results are averaged over 3 random seeds.

\subsection{Main Results: Unsupervised Transfer}

Table~\ref{tab:main_results} reports RMSE and MAE on the target
platform C under unsupervised adaptation ($k=0$).

\begin{table}[!t]
  \caption{Unsupervised Cross-Platform Transfer to Platform C
           ($k=0$ Target Labels; $\downarrow$ Lower is Better)}
  \label{tab:main_results}
  \centering
  \renewcommand{\arraystretch}{1.2}
  \begin{tabular}{lcc}
    \toprule
    \textbf{Model} & \textbf{RMSE} $\downarrow$ & \textbf{MAE} $\downarrow$ \\
    \midrule
    Source-Only (A only)           & 0.79$\pm$0.01 & 0.62$\pm$0.01 \\
    Pool-NoAdapt (A+B)             & 0.76$\pm$0.01 & 0.60$\pm$0.01 \\
    DANN-Only \cite{ganin2017dann} & 0.70$\pm$0.01 & 0.55$\pm$0.01 \\
    \midrule
    \textbf{ADAPT-MS (Ours)}       & \textbf{0.66$\pm$0.01} & \textbf{0.52$\pm$0.01} \\
    \bottomrule
  \end{tabular}
\end{table}

ADAPT-MS achieves a 16\% relative RMSE reduction over Pool-NoAdapt
and a 5.7\% reduction over DANN-Only:
\begin{equation}
  \Delta_{\mathrm{RMSE}} = \frac{0.70 - 0.66}{0.70} \approx 5.7\%.
  \label{eq:improvement}
\end{equation}
The gap between DANN-Only and ADAPT-MS is primarily attributed to
the calibration layer correcting the systematic over-prediction bias
caused by Platform C's inflated 5-star usage (mean rating 4.70
vs.\ 4.10 on B).

\subsection{Few-Shot Adaptation Curve}

Table~\ref{tab:fewshot} reports RMSE as the number of labeled
target samples $k$ increases, comparing fine-tuning all parameters
against updating only the calibration layer and gate.

\begin{table}[!t]
  \caption{Few-Shot Adaptation Performance (RMSE on Platform C;
           $\downarrow$ Lower is Better)}
  \label{tab:fewshot}
  \centering
  \renewcommand{\arraystretch}{1.2}
  \begin{tabular}{lcccc}
    \toprule
    \textbf{Model}
      & $k=0$ & $k=50$ & $k=200$ & $k=1000$ \\
    \midrule
    Fine-Tune (all params)
      & 0.79 & 0.73 & 0.69 & 0.66 \\
    \textbf{ADAPT-MS (calib+gate only)}
      & \textbf{0.66} & \textbf{0.63} & \textbf{0.61} & \textbf{0.60} \\
    \bottomrule
  \end{tabular}
\end{table}

Two observations stand out. First, at $k=0$ ADAPT-MS already
achieves 0.66, whereas full fine-tuning collapses to 0.79 (the
source-only baseline) without any target labels. Second, the
gap narrows as $k$ increases, with both methods converging near
0.60--0.66 at $k=1000$. This confirms that updating only the
lightweight calibration parameters is both data-efficient and
stable.

\subsection{Sensitivity to Alignment Weight}

Table~\ref{tab:lambda} shows target-platform RMSE as a function
of the domain alignment weight $\lambda$.

\begin{table}[!t]
  \caption{Sensitivity to Domain Alignment Weight $\lambda$
           (RMSE on Platform C; $\downarrow$ Lower is Better)}
  \label{tab:lambda}
  \centering
  \renewcommand{\arraystretch}{1.2}
  \begin{tabular}{lcccc}
    \toprule
    $\lambda$ & 0.0 & 0.1 & 0.5 & 1.0 \\
    \midrule
    Target RMSE & 0.76 & 0.70 & \textbf{0.66} & 0.68 \\
    \bottomrule
  \end{tabular}
\end{table}

Performance peaks at $\lambda=0.5$: too little alignment
($\lambda\!=\!0.1$) fails to remove platform-specific features,
while too strong alignment ($\lambda\!=\!1.0$) suppresses
predictive signals that are common across platforms but
superficially correlated with platform identity.

\subsection{Ablation Study}

Table~\ref{tab:ablation} isolates the contribution of each
ADAPT-MS component on Platform C under unsupervised transfer.

\begin{table}[!t]
  \caption{Ablation Study on Platform C (RMSE, $k=0$;
           $\downarrow$ Lower is Better)}
  \label{tab:ablation}
  \centering
  \renewcommand{\arraystretch}{1.2}
  \begin{tabular}{lc}
    \toprule
    \textbf{Model Variant} & \textbf{RMSE} $\downarrow$ \\
    \midrule
    Full ADAPT-MS                          & \textbf{0.66} \\
    \quad w/o calibration layer            & 0.71          \\
    \quad w/o domain-adversarial alignment & 0.74          \\
    \quad w/o modality dropout / gating    & 0.70          \\
    \quad w/o canonical behavior features  & 0.73          \\
    \midrule
    Text-only (no behavior, no alignment)  & 0.76          \\
    \bottomrule
  \end{tabular}
\end{table}

The calibration layer and domain-adversarial alignment provide
the two largest contributions (RMSE $+$0.05 and $+$0.08
respectively when removed). Modality dropout and gating are
particularly important for Platform C, where 40\% of instances
lack behavioral logs: removing these components increases RMSE
by 0.04.

\subsection{Pairwise Transfer Analysis}

Table~\ref{tab:pairwise} reports pairwise transfer results between
all three platform pairs, confirming that ADAPT-MS consistently
outperforms the no-adaptation baseline regardless of transfer direction.

\begin{table*}[!t]
  \caption{Pairwise Cross-Platform Transfer RMSE
           ($\downarrow$ Lower is Better; Pool-NoAdapt vs.\ ADAPT-MS)}
  \label{tab:pairwise}
  \centering
  \renewcommand{\arraystretch}{1.2}
  \begin{tabular}{lcccccc}
    \toprule
    & \multicolumn{2}{c}{\textbf{A$\to$B}}
    & \multicolumn{2}{c}{\textbf{B$\to$C}}
    & \multicolumn{2}{c}{\textbf{A+B$\to$C}} \\
    \cmidrule(lr){2-3}\cmidrule(lr){4-5}\cmidrule(lr){6-7}
    \textbf{Model}
      & RMSE & MAE & RMSE & MAE & RMSE & MAE \\
    \midrule
    Pool-NoAdapt
      & 0.74 & 0.58 & 0.78 & 0.61 & 0.76 & 0.60 \\
    \textbf{ADAPT-MS}
      & \textbf{0.69} & \textbf{0.54}
      & \textbf{0.70} & \textbf{0.55}
      & \textbf{0.66} & \textbf{0.52} \\
    \midrule
    Relative gain & 6.8\% & 6.9\% & 10.3\% & 9.8\% & 13.2\% & 13.3\% \\
    \bottomrule
  \end{tabular}
\end{table*}

The largest gain (13.2\% RMSE reduction) is observed on
A+B$\to$C, where both distribution shift and rating scale
shift are most pronounced. The A$\to$B transfer shows
smaller gains because the two platforms share a more similar
learner population and rating culture.

\subsection{Failure Mode Analysis}

We identify two primary failure modes in cross-platform transfer:
\begin{enumerate}
  \item \emph{Logging schema gaps}: Platform C lacks fine-grained
    video interaction events (seek/rewind), making the rewatch
    feature $\rho_i$ uncomputable. Gated fusion mitigates but
    does not fully eliminate this gap.
  \item \emph{Domain-specific vocabulary}: Platform-specific UI
    terminology (e.g., ``XP points'', ``nanodegree'') causes
    spurious correlations in the LLM representations that
    domain-adversarial training struggles to fully remove,
    motivating lightweight target-domain language adaptation
    as future work.
\end{enumerate}

\section{Conclusion}
\label{sec:conclusion}

We proposed ADAPT-MS, a cross-platform domain adaptation
framework for multi-modal MOOC learner satisfaction prediction.
By combining domain-adversarial representation alignment, a
platform-specific affine calibration layer, and missing-modality-
robust gated fusion with modality dropout, ADAPT-MS achieves
target-platform RMSE of 0.66 in the zero-label setting and 0.60
with 1000 labeled target samples---outperforming naive pooling,
source-only, and alignment-only baselines. Ablation studies confirm
that all four components contribute independently, with calibration
and adversarial alignment providing the largest gains. The few-shot
adaptation curve demonstrates that ADAPT-MS remains effective with
as few as 50 labeled target samples, making it practical for real-
world platform onboarding scenarios.

Future directions include: (i) extending to multilingual
cross-platform settings where linguistic style shift is compounded
by language shift, (ii) handling temporal drift within a single
platform (concept drift over cohorts), (iii) exploring online
calibration updates using streaming target data, and (iv)
connecting cross-platform adaptation to fairness auditing to ensure
that calibration does not disproportionately harm underrepresented
learner groups on the target platform.

\appendix
\section{Reproducibility Details}
\label{sec:repro}

\subsection{Data Splits and Leakage Control}
Within each platform, course runs are sorted chronologically.
Training, validation, and test sets use the earliest 70\%, next
15\%, and most recent 15\% of runs, respectively. For pairwise
transfer experiments, the source platform training set is the
full chronological 85\% split; the target test set is its most
recent 15\%.

\subsection{Canonical Feature Computation}
All six canonical behavioral features (Eq.~\eqref{eq:behavior})
are computed within a fixed 28-day window from course start.
Platform-specific features not in the canonical set are excluded.
Missing features are imputed with the within-platform training
mean and flagged with the modality indicator $m_i=0$.

\subsection{Hyperparameter Search}
For each model variant, we run random search with 20 trials on
the validation split of the source platforms. Tuned parameters:
learning rate $\in\{10^{-4}, 3\times10^{-4}, 10^{-3}\}$,
dropout $\in\{0.1,0.2,0.3\}$, fusion hidden size $\in\{256,512\}$,
and alignment weight $\lambda\in\{0.0,0.1,0.3,0.5,1.0\}$.
The same budget is used for all variants.

\subsection{Computational Budget}
ADAPT-MS trains in approximately 1.5 hours per fold on a single
NVIDIA A100 GPU (40~GB). Text embeddings are pre-computed and
cached (approximately 30 minutes per platform). Calibration-only
few-shot adaptation requires less than 5 minutes per $k$ setting.

\bibliographystyle{IEEEtran}
\bibliography{references}

\end{document}